\documentclass[pra,nofootinbib]{revtex4}

\usepackage{graphicx}
\usepackage{amssymb}
\usepackage{amsmath}
\usepackage{amsfonts}
\usepackage{epstopdf}
\usepackage{epsfig}
\usepackage{wrapfig}

\begin{document}
\title{Model of quantum measurement and thermodynamical cost of accuracy and stability of information processing}

\author{Robert Alicki,}
\affiliation{Institute of Theoretical Physics and Astrophysics, University of
Gda\'nsk, Poland }

\begin{abstract}
The quantum measurement problem is revisited and discussed in terms of a new solvable measurement model which basic ingredient is the quantum model of a controlled single-bit memory. The structure of this model involving strongly coupled spin and quantum harmonic oscillator allows to  define stable pointer states as well-separated Gaussian states of the quantum oscillator and analyze the transition from quantum to classical regime. The relations between accuracy of measurement, stability of pointer states, effective temperature of joint thermal and quantum noise and minimal work needed to perform the bit-flip are derived. They differ from those based on the Landauer principle and are used to analyze thermodynamic efficiency of quantum Szilard engine and imply more realistic estimations of minimal amount of work needed to perform long computations.

\end{abstract} 
\maketitle

\section{Introduction}

The \emph{problem of quantum measurement} is often considered as the fundamental unsolved problem of quantum physics which either belongs rather to philosophy of science or for its solution one needs new inputs which should strongly modify the present paradigm. On the other hand a number of ideas which appeared in few recent decades centered around the notions of decoherence and \emph{emergence of classical world} \cite{Joos:2003},  \emph{stable pointer states} \cite{Zurek:2003}, \emph{continuous measurements} and \emph{stochastic unraveling} \cite{Barchielli:2010}, seem to give quite plausible insights into the measurement problem. The rigorous mathematical scheme for this framework is provided by the quantum theory of open systems \cite{Breurer:2002}, in particular by the theory of completely positive quantum dynamical semigroups governed by quantum Markovian Master equations obtained in the weak coupling regime \cite{Davies:1974, Alicki:2006}. The recent review article on quantum measurement theory \cite{Allah:2012} contains an extensive list of references and present an example of measurement device based on Curie-Weiss model. Another aspect of measurement theory is the thermodynamical cost of measurement and its information-theoretical meaning. Starting with the classical Szilard argument the discussion led to the formulation of Landauer principle \cite{Landauer:1961} and its interpretation by Bennett \cite{Bennett:2003} which can be summarized by the statement that any irreversible operation on an information carrier immersed in a  thermal equilibrium environment at the temperature $T$ and which lead to an erasure of a single bit cost at least $T\ln 2$ of work (I put always $k_B = \hbar =1$). Further developments  based, essentially, on the energy-entropy balance can be found in \cite{Sagawa:2012}.
\par
The approach presented in this paper incorporates, to some extend, those ideas, but on the other hand introduces new ones motivated by the notions of measurement error and stability of information processing which characterize the \emph{quality} of those processes. This allows to find quite deep connections between the issues mentioned above and leads to  new expressions for the thermodynamical cost of information processing and efficiency of the quantum Szilard engine which are different than those based on the Landauer formula. The model itself is also different, the semiclassical pointer is permanently and strongly coupled to the quantum "interface" spin-1/2 system and the later is directly coupled to the observed quantum system. The source of work needed to perform a measurement (or logical CNOT gate) is explicitly introduced.
\par
Similarly to the arguments used by Szilard and the followers I discuss in details a particular  quantum model of a measuring device which at the same time describes a quantum model of a single bit memory. I believe that simplicity and universality of this model allow to draw general conclusions and derive the general bounds of the thermodynamical character. In contrast to the original Szilard model and its later modifications the presented model is fully quantum and derivable in a rigorous way from the first principle Hamiltonian theory. The model possess also all (in)stability properties which are more or less explicitly assumed in the quantum measurement theory.
\par
In order to avoid the problem of distinguishability of pointer states - \emph{the observation of the pointer requires another measuring instrument, which in turn requires yet another instrument, and so on, in such a way that the whole process involves an infinite regression ending up in the observer's brain} \cite{Sewell:2007} - I assume that :\\

\emph{There exist stable pointer states which can be distinguished with an error probability $\epsilon$ given by their overlap quantified by the quantum transition probability. The life-time of the pointer states scales like $\frac{1}{\epsilon}$ and the thermodynamical cost of pointer states recognition vanishes with $\epsilon\to 0$.}\\

\par
The model system consists of a harmonic oscillator strongly coupled to a spin-$1/2$ and weakly interacting  with a heat bath. The harmonic oscillator part represents the pointer of the measuring device with two (zero-momentum) coherent states separated by a dimensionless "distance" $2D$ and serving as two pointer positions. Those states correspond to two degenerated ground states of the device: "spin up , oscillator localized at $x = D$" or  "spin down, oscillator localized at $x = -D$". The overlap of two coherent states  given by $\epsilon = e^{-4D^2}$ yields  the probability of error in the process of distinguishing pointer states at zero temperature. I first demonstrate these properties for the zero temperature environment (Section II) and then for the finite temperature case (Section III). Obviously, for finite temperatures the pointer states are not anymore pure states but they remain well-localized Gaussian states with desired stability. For all temperatures  the averaged minimal amount of work supplied by the time-dependent Hamiltonian is given by $\bar{W} = \frac{1}{2}{\omega_0} (2D)^2$ ($\omega_0$- frequency of the oscillator). On the other hand the formula for the measurement error is given by $\epsilon = e^{-\bar{W}/\Theta}$, where $\Theta = \Theta[T,\omega_0]$ is an average quantum oscillator energy (including "zero-point energy") at the temperature $T$ and can be treated as the effective "noise temperature" characterizing both, thermal and quantum fluctuations on the same footing. The same Boltzmann-like factor characterizes stability of the pointer states, i.e. in the leading order of magnitude the dissipative tunneling rate between two pointer states is proportional to $ e^{-\bar{W}/\Theta}$
\par

This picture agrees with the standard intuition that the measuring apparatus should contain a "classical part" with clearly distinguishable stable pointer states. Those states are localized non-orthogonal and generally mixed semiclassical states. One is mainly interested in the semiclassical, low $\epsilon$, domain and therefore \emph{recognition cost} can be neglected in comparison with other ones. The pointer states  display  also \emph{cat states} behavior, i.e. their superpositions  decay  quickly with the decay rate $\sim D^2$. 
\par
As one can expect the more stable are the pointer states or equivalently the  more accurate is the measurement, the more work must be invested to change a pointer state or equivalently to record the measurement result.
The obtained formulas give  more realistic estimations of the thermodynamic cost of the measurement and information processing than the lowest bound given by Landauer formula which follows from the II-law of thermodynamics and does not depend on the quality of the measurement and recording processes. This is heuristically obvious because more accurate measurements and information processing on more robust information carriers are always more expensive in any sense of this word.
\par
In Section IV a quantum version of the Szilard engine is revisited. In contrast to the standard approach the accuracy of the measurement which powers the engine is taken into account and the total balance of energy is computed using the introduced measurement model. It allows to compute the universal bound on the efficiency of such an engine defined by the ratio of extracted work to work invested in the measurement. Concluding remarks address the issues of minimal thermodynamical cost of long computations and the fundamental conflict between stability and reversibility of information processing.
\section{A quantum model of a stable information carrier, zero temperature case}
The basic ingredients of the spin-oscillator system (SOS) and its dynamical features are illustrated on Fig.1 and Fig.2.  The model consists of a spin-$1/2$ described by the standard Pauli matrices $\hat{\sigma}^k , k=1,2,3,\pm$ and the harmonic oscillator with  the canonical operators $\hat{a}, \hat{a}^{\dagger}$. The SOS Hamiltonian reads
\begin{equation}
\hat{H} = \omega_0 (\hat{a}^{\dagger} - D\hat{\sigma}^3)(\hat{a} -D\hat{\sigma}^3)= \omega_0 \hat{a}^{\dagger}\hat{a} - D(\hat{a}^{\dagger}+\hat{a})\hat{\sigma}^3 + \omega_0 D^2 ,\quad  \omega_0 , D > 0 ,
\label{ham_TLSQB}
\end{equation}
where, \eqref{ham_TLSQB} is a physical renormalized Hamiltonian which includes the lowest order corrections due to the interaction with an environment. The following notation for spin states, oscillator coherent states and joint spin-oscillator states is used
\begin{equation}
\hat{\sigma}^3 |\pm\rangle = \pm |\pm\rangle, \quad \hat{a}|\alpha\rangle = \alpha |\alpha\rangle, \quad \alpha\in\mathbf{Z},\quad |\mu;\alpha\rangle \equiv |\mu\rangle|\alpha\rangle, \quad \mu= \pm.
\label{coherent}
\end{equation}
The analysis of SOS is simplified by using a new set of canonical operators $ \hat{b}, \hat{b}^{\dagger}$ and new set of Pauli matrices $\hat{\tau}^{k}$ obtained  from $\hat{a}, \hat{a}^{\dagger} , \hat{\sigma}^k $ by the following  unitary \emph{dressing transformation}
\begin{equation}
\hat{a} \mapsto \hat{b} = \hat{U}^{\dagger} \hat{a} \hat{U},\quad \hat{\sigma}^k \mapsto \hat{\tau}^{k} = U^{\dagger} \sigma^k U\ ,\  U=e^{D( \hat{a}- \hat{a}^{\dagger})\hat{\sigma}^3}.
\label{dress}
\end{equation}
Notice, that under this transformation 
\begin{equation}
\hat{\tau}^3 = \hat{\sigma}^3, \quad \hat{b}= \hat{a}-  D\sigma^3, ,
\label{dress1}
\end{equation}
and therefore the SOS Hamiltonian \eqref{ham_TLSQB}  can be written as
\begin{equation}
\hat{H}=\omega_0 \hat{b}^{\dagger}\hat{b} - \omega_0 D^2 .
\label{ham_TLSQB_1}
\end{equation}
The dressing transformation can be expressed in terms of new variables
\begin{equation}
\hat{U} = e^{D( \hat{a}- \hat{a}^{\dagger})\hat{\sigma}^{3}}=e^{D( \hat{b}- \hat{b}^{\dagger})\hat{\tau}^{3}}.
\label{dress1}
\end{equation}
From the form \eqref{ham_TLSQB_1} of the SOS Hamiltonian it follows that its ground state satisfies $ \hat{b}|\psi\rangle = 0$ and hence, due to \eqref{dress1} is double degenerated and given by the product of a spin $\hat{\sigma}^3$-eigenstate
and the corresponding harmonic oscillator coherent state $|\pm D\rangle $. The overlap ( transition probability ) between those coherent states is given by
\begin{equation}
\epsilon = |\langle D |- D\rangle|^2 = e^{-4D^2} .
\label{trans_0}
\end{equation}
One should notice that only for $\epsilon \in [0, 1/2)$ those states are distinguishable with error probability $\epsilon$ and can serve as pointer states.
\par
The ground states and the excited ones obtained by a spin-flip play an important role and deserve a short-hand notation
\begin{equation}
|\pm ;\pm D\rangle \equiv |\Omega_{\pm}\rangle , \quad |\pm ;\mp D\rangle \equiv |\Omega^*_{\pm}\rangle .
\label{ground}
\end{equation}
Notice, that the sign $\pm$ in $|\Omega_{\pm}\rangle$ and $|\Omega^*_{\pm}\rangle $ is determined by a spin state.
\par
Assume that SOS is weakly coupled to a large quantum system at zero temperature by means of the interaction Hamiltonian which can be decomposed into three independent terms written in original and dresses variables 
\begin{eqnarray}
\label{Ham_int_1}
\hat{H}^{(o)}_{int} &=& (\hat{a} + \hat{a}^{\dagger}) \hat{F_o} = (\hat{b} + \hat{b}^{\dagger} - 2D\hat{\tau}^3) \hat{F}_o,\\
\label{Ham_int_2}
\hat{H}^{(3)}_{int} &=& \hat{\sigma}^3 \hat{F}_3 = \hat{\tau}^3 \hat{F}_3,
 \\
\hat{H}^{(1)}_{int} &=& \hat{\sigma}^1 \hat{F}_1 = \bigl(e^{2D( \hat{b}- \hat{b}^{\dagger})}\hat{\tau}^{+} + e^{-2D( \hat{b}- \hat{b}^{\dagger})}\hat{\tau}^{-}\bigr) \hat{F}_1
\label{Ham_int_3}
\end{eqnarray}
where $F_{j}$ are independent environment observables and $\hat{\tau}^{\pm}= (\hat{\tau}^{1}\pm i\hat{\tau}^{2})/2$. The relevant properties of the environment are encoded in the Fourier transforms of the autocorrelation functions 
\begin{equation}
G_j(\omega)= \int_{-\infty}^{+\infty} e^{i\omega t}\langle \hat{F}_j(t)\hat{F}_j\rangle_{0} dt \geq 0 ,
\label{spectral_bath}
\end{equation}
where $\hat{F}_j(t)$ evolves according to the environment Hamiltonian, $\langle \cdot \rangle_{0}$ denotes the average with respect to zero temperature (ground) state of the environment what implies  that $G_j(\omega) = 0$ for $\omega < 0$. Another ingredient of the construction are the Fourier components of the SOS operators which appears in \eqref{Ham_int_1}, \eqref{Ham_int_2}, \eqref{Ham_int_3} and evolve in the Heisenberg picture according to the SOS Hamiltonian
\begin{eqnarray}
\label{Fourier_1}
e^{i\hat{H} t}(\hat{a} + \hat{a}^{\dagger})e^{-i\hat{H} t}  &=& e^{-i\omega_0 t}\hat{b} + e^{i\omega_0 t}\hat{b}^{\dagger} - 2D\hat{\tau}^3,
\label{Fourier_2}\\
e^{i\hat{H} t} \hat{\sigma}^3 e^{-i\hat{H} t}  &=& \hat{\sigma}^3 \equiv \hat{\tau}^3\\
\label{Fourier_3}
e^{i\hat{H} t} \hat{\sigma}^1 e^{-i\hat{H} t}  &=&  e^{2D( e^{-i\omega_0 t}\hat{b}- e^{i\omega_0 t}\hat{b}^{\dagger})}\hat{\tau}^{+} + e^{-2D( e^{-i\omega_0 t}\hat{b}- e^{i\omega_0 t}\hat{b}^{\dagger})}\hat{\tau}^{-}.
\end{eqnarray}
Under standard assumptions concerning the ergodic properties of the environment and the product state assumption for the SOS-environment state one can derive the Schroedinger picture, quantum Markovian master equation for the density matrix of SOS valid in the weak coupling regime \cite{Davies:1974}.
\subsection{SOS irreversible dynamics without tunneling}

 For the clarity of the presentation I omit first the contribution from the incoherent tunneling described by \eqref{Ham_int_3}, \eqref{Fourier_3} and obtain the simplified master equation
\begin{equation}
\frac{d\hat{\rho}}{dt} = -i[\hat{H} , \hat{\rho}] + \frac{1}{2}\gamma\bigl([\hat{b}, \hat{\rho}\hat{b}^{\dagger}] + [\hat{b} \hat{\rho}, \hat{b}^{\dagger}]\bigr) -\frac{1}{2}\Gamma [\hat{\tau}^3,[\hat{\tau}^3, \hat{\rho}]] .
\label{MME}
\end{equation}
where the dissipation rate $\gamma = G_o(\omega_0)$, and the pure decoherence rate $\Gamma = 4D^2 G_o(0) + G_3(0)$.
\par
In order to study the dynamics given by \eqref{MME} it is useful to compute explicitly the evolution of the following rank-1 operators, constructed from the spin eigenstates and coherent states of the oscillator, which can be used as building blocks for an arbitrary initial density matrix
\begin{equation}
|\mu ;\alpha \rangle\langle\beta ;{\nu}|\mapsto e^{-\frac{\Gamma}{2}(\mu -\nu)^2 t} e^{\Phi(t)}|\mu ;\alpha_{\mu}(t)\rangle\langle\beta_{\nu}(t);\nu|
\label{rank_1}
\end{equation}
where for $\mu=\pm$
\begin{equation}
\alpha_{\mu}(t) = \alpha e^{-(i\omega_0 + \frac{\gamma}{2})t} + \mu D \bigl[ 1- e^{-(i\omega_0 + \frac{\gamma}{2})t}\bigr]
\label{alpha}
\end{equation}
and
\begin{equation}
\Phi(t) = \bigl(e^{-\gamma t}-1\bigr) \bigl[\frac{1}{2}(\alpha -\mu D)^2 + \frac{1}{2}(\beta -\nu D)^2  - (\alpha -\mu D)(\overline{\beta} -\nu D)\bigr].
\label{Phi}
\end{equation}
The formulas of above can be proved by inserting the solution \eqref{rank_1}, \eqref{alpha}, \eqref{Phi} into master equation \eqref{MME} and using the following identity valid for any time-dependent coherent state
\begin{equation}
\frac{d}{dt}|\alpha(t)\rangle = -|\alpha (t)|\frac{d|\alpha (t)|}{dt}|\alpha(t)\rangle + \frac{d\alpha(t)}{dt}\hat{a}^{\dagger}|\alpha(t)\rangle .
\label{der_alpha}
\end{equation}
\par
It follows from the above  solution that any mixture of ground states $p_{-}|\Omega_{-}|\rangle\langle \Omega_{-}| +  p_{+}|\Omega_{+}|\rangle\langle \Omega{+}|$ is a stationary state and any initial state $\hat{\rho}$ tends asymptotically to such stationary state with $p_{\pm}$ determined by the initial spin populations
\begin{equation}
p_{\pm}= \frac{1}{2}\mathrm{Tr}\bigl(\hat{\rho}(\hat{I} \pm \hat{\sigma}^3) \bigr). 
\label{prob_in}
\end{equation}
Writing a state $\hat{\rho}$
as a $2\times 2$ matrix $[\hat{\rho}_{\mu\nu}]$ one can notice that  decay of the off-diagonal elements $[\hat{\rho}_{+ -}]=[\hat{\rho}^{\dagger}_{-+}]$ is determined by the decoherence rate $\Gamma$ which in the classical regime  ($D >> 1$) scales like $\sim D^2$. For example the initial superposition of spin states and fixed coherent state  $c_{-}|-;\alpha \rangle + c_{+}|+;\alpha \rangle $ evolves into a mixed state  with  decaying off-diagonal quantum coherences. This decay is fast in the classical regime ($D >> 1$) what can be seen from the explicit solution
\begin{eqnarray}
\nonumber
\hat{\rho}(t) &=& |c_{-}|^2|-;\alpha_{-}(t)\rangle\langle\alpha_{-}(t);-| + |c_{+}|^2|+;\alpha_{+}(t)\rangle\langle\alpha_{+}(t);+| \\
\label{state_1}
&+& \Bigl[\exp \Bigl\{-\frac{1}{2}\bigl(1- e^{-\gamma t}\bigr)( D^2 +i\zeta) - 2\Gamma t\Bigr\} c_{-}\overline{c_{+}}\,|-;\alpha_{-}(t)\rangle\langle\alpha_{+}(t); +| + \mathrm{h.c.}\Bigr] .
\end{eqnarray}
where $\zeta$ denotes an irrelevant phase.
\par
For spin-diagonal components one observes structural stability of  coherent states $|\alpha_{\pm}(t)\rangle$ which evolve along the classical trajectory
in phase space towards the attractor $|\Omega_{\pm}\rangle$. On the other hand the superposition of coherent states  with a fixed spin state (say $|-\rangle$) $c_{1}|-;\alpha \rangle + c_{2}|-;\beta \rangle $
evolves into a mixed state
\begin{eqnarray}
\nonumber
\hat{\rho}(t) &=& |c_{1}|^2|-;\alpha_{-}(t)\rangle\langle\alpha_{-}(t);-| + |c_{2}|^2 |-;\beta_{-}(t)\rangle\langle\beta_{-}(t);-| \\
\label{state_2}
&+& \Bigl[\exp \Bigl\{-\frac{1}{2}\bigl(1- e^{-\gamma t}\bigr)(|\alpha -\beta|^2 +i \zeta )\Bigr\} c_{1}\overline{c_{2}}\, |-;\alpha_{-}(t)\rangle\langle\beta_{-}(t);-| + \mathrm{h.c.}\Bigr]
\end{eqnarray} 
where $\zeta$ is again an irrelevant phase. The interference terms between two coherent states decays rapidly for macroscopically distinguishable case , i.e. for $|\alpha -\beta|^2 >> 1$. The computations of above illustrate the expected emerging classical properties of the coherent SOS states $|\alpha ,{\pm}\rangle$ which are necessary for the appearance of well-determined, distinguishable (up to the error probability $\epsilon = e^{-4D^2}$), and robust pointer states.

\begin{figure}[tb]
    \centering
    \includegraphics[width=0.8\textwidth,angle=270]{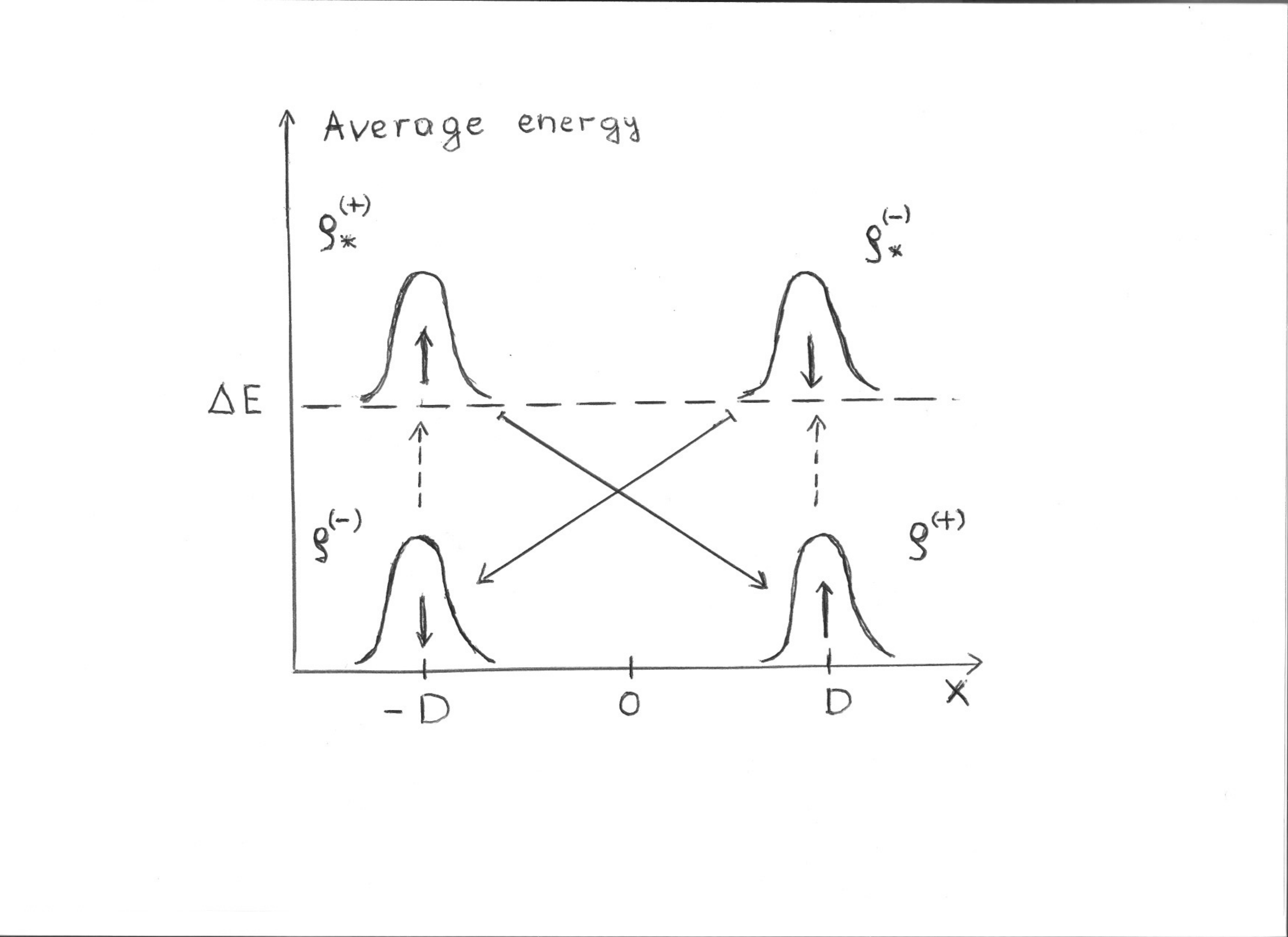}
    \caption{Stable SOS states $\rho^{(\pm)}$ and their excitations $\rho^{(\pm)}_*$ (or $|\Omega_{\pm}\rangle$ and, $|{\Omega^*}_{\pm}\rangle$ for zero temperature case). Gaussians depict localized pointer states with arrows inside corresponding to spin states.  The long solid arrows represent dissipation routes, while the dashed ones first steps of tunneling process.}
    \label{nazwa}
 \end{figure}

\subsection{Stability with respect to tunneling}
One can expect that the process of environmentally driven tunneling between spin states of SOS described by the interaction Hamiltonian \eqref{Ham_int_3} is suppressed due to the energy barrier  $\Delta E= 4D^2\omega_0$ between the ground states $|\Omega_{\pm}\rangle$ and the corresponding spin-flipped states $|\Omega^*_{\pm}\rangle$. Therefore, the tunneling rate should scale exponentially  $\Gamma_{tun} \sim e^{-\Delta E/\omega_0} = e^{-4D^2}$. A more rigorous argument can be obtained using an additional term in the Master equation for SOS. For simplicity we assume that the spectral density of the reservoir associated with \eqref{Ham_int_3} vanishes for $\omega \geq \omega_{cut}< \omega_0$.
Therefore, only the Fourier component of \eqref{Fourier_3} corresponding to zero Bohr frequency and given by
\begin{equation}
\hat{B}_0 ={\hat{B}_0}^{\dagger} = e^{-2D^2}\sum_{n=0}^{\infty} (-1)^n \frac{(2D)^{2n}}{(n!)^2} \bigl(\hat{b}^{\dagger}\bigr)^n \hat{b}^n
\label{W0}
\end{equation}
enters the extended master equation. The corresponding pure decoherence term in the quantum dynamical semigroup generator reads
\begin{equation}
\mathcal{L}^{(1)}\hat{\rho} = = -\frac{1}{2}G_{1}(0)[\hat{B}_0 \hat{\tau}^1,[\hat{B}_0 \hat{\tau}^1, \hat{\rho}]] 
\label{gen_new}
\end{equation}
The tunneling rate can be rigorously defined as the initial probability flow  from the spin state $|-\rangle$ to the spin state $|+\rangle$ for the SOS initially in the  ground state $|\Omega_{-}\rangle$
\begin{equation}
\Gamma_{tun} = \frac{1}{2}\mathrm{Tr}\Bigl(\hat{\tau}^3 \frac{d\hat{\rho}}{dt}|_{t=0} \Bigr)=\frac{1}{2}\mathrm{Tr}\Bigl(\hat{\tau}^3\mathcal{L}^{(1)}(|\Omega_{-}\rangle\langle \Omega_{-}|) \Bigr)= \frac{1}{2}G_1(0)\langle \Omega_{-}|(\hat{B}_0)^2 |\Omega_{-}\rangle
\label{tunrate}
\end{equation}
Notice that the transition between spin states is only due to the tunneling term \eqref{gen_new}. Using  \eqref{W0} and the relation $\hat{b}|\Omega_{\pm}\rangle =0$ one obtains the final expression
\begin{equation}
\Gamma_{tun} = \frac{1}{2}G_1(0) e^{-4D^2}
\label{tunrate1}
\end{equation}
with the same scaling as that obtained using heuristic reasoning. The parameter $G_1(0)$ is the pure decoherence rate for the spin decoupled from the oscillator, while the exponential factor describes the stabilizing effect of spin-oscillator coupling. A more detailed discussion of the tunneling process, for the finite temperature case, is presented in the Appendix.

\begin{figure}[tb]
    \centering
    \includegraphics[width=0.8\textwidth,angle=270]{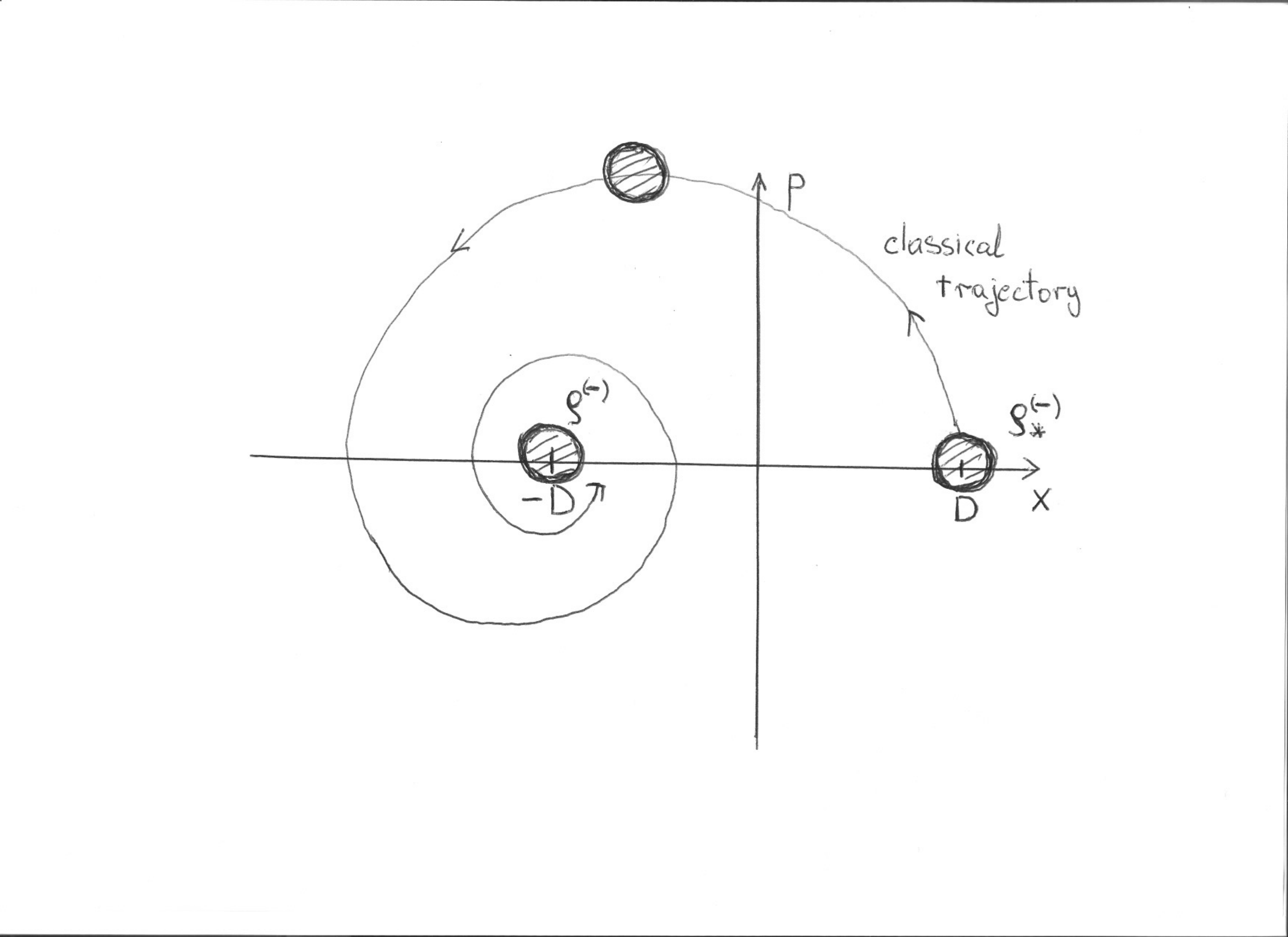}
    \caption{Phase-space picture of the recording process. Stable SOS state $\rho^{(+)}$ is excited to the state $\rho^{(-)}_*$ and then evolves along the damped harmonic oscillator classical trajectory towards the final stable state SOS $\rho^{(-)}$}
    \label{nazwa}
 \end{figure}

\subsection{Quantum measurement model}
The SOS coupled to the zero-temperature bath as described by the eq, \eqref{MME} can be used as a model of quantum measurement. The measured dichotomic observable of the observed system $\mathcal{O}$ has a structure $\hat{X} = \hat{P}_{+} - \hat{P}_{-}$ with two orthogonal projectors satisfying $ \hat{P}_{-} + \hat{P}_{+} = I$. The measurement of $\hat{X}$
is performed by the coupling of $\mathcal{O}$ and SOS by means of the following time-dependent Hamiltonian
\begin{equation}
\hat{H}_{M} = f(t)\hat{\sigma}^{1}\hat{P}_{-} 
\label{ham_measure}
\end{equation}
where $f(t)$ is assumed to be a fast pulse of the duration $t_M$ concentrated around $t=0$ and satisfying $\int_{-\infty}^{\infty} f(t) dt = \pi$. This Hamiltonian executes a CNOT gate on the spin controlled by the value of the observable $\hat{X}$. On the average, in half of the cases the spin flip is performed and the energy of SOS is increased by $\Delta E = 4D^2\omega_0 $. Therefore the averaged work performed by the pulse is given by
\begin{equation}
\bar{W} = \frac{1}{2}\Delta E = 2D^2\omega_0 . 
\label{work_0}
\end{equation}
The initial state of the SOS is one of the ground states, say $|\Omega_{+}\rangle$ with $D >>1$, while the initial state of $\mathcal{O}$ is of the form
\begin{equation}
|\phi_{\mathcal{O}}(0)\rangle = c_{-}|\phi_{-}\rangle + c_{+}|\phi_{+}\rangle ,\quad \hat{P}_{-}|\phi_{-}\rangle =|\phi_{-}\rangle, \quad  \hat{P}_{+}|\phi_{+}\rangle = |\phi_{+}\rangle .
\label{O_state}
\end{equation}
If the measurement time $t_M$ is much shorter than the other dynamical time-scales the state of SOS $+ \mathcal{O}$
after measurement is given by
\begin{equation}
|\Phi_{tot}(t_M)\rangle = c_{-}|\phi_{-}\rangle|\Omega^*_{-}\rangle + c_{+}|\phi_{+}\rangle |\Omega_{+}\rangle = \bigl(c_{-}|\phi_{-}\rangle|-\rangle + c_{+}|\phi_{+}\rangle |+\rangle \bigr) |D\rangle.
\label{tot_state}
\end{equation}
Notice that the state of $\mathcal{O}$ after the measurement is given by the following reduced density matrix
\begin{equation}
\hat{\rho}_{\mathcal{O}}(t_M)= |c_{-}|^2|\phi_{-}\rangle\langle \phi_{-}| + |c_{+}|^2|\phi_{+}\rangle\langle \phi_{+}| .
\label{Ostate_post}
\end{equation}
For $t > t_M$ SOS and $\mathcal{O}$ evolve independently and according to \eqref{state_1}, if the tunneling is neglected, the reduced density matrix of SOS reads
\begin{eqnarray}
\nonumber
\hat{\rho}(t) &=& |c_{-}|^2|\Omega^*_{-}(t)\rangle\langle \Omega^*_{-}(t)| + |c_{+}|^2|\Omega_{+}\rangle\langle\Omega_{+}| \\
\label{SOS_state}
&+& \Bigl[\exp \Bigl\{-\frac{1}{2}\bigl(1- e^{-\gamma t}\bigr) (D^2 + i\zeta) - 2\Gamma t\Bigr\} c_{-}\overline{c_{+}}\,|\Omega^*_{-}(t)\rangle\langle\Omega_{+}| + \mathrm{h.c.}\Bigr] .
\end{eqnarray}
where
\begin{equation}
|\Omega^*_{-}(t)\rangle \equiv |-;D_{-}(t)\rangle,\quad D_{-}(t) = D \bigl(2 e^{-(i\omega_0 +\frac{\gamma}{2})t}-1 \bigr) ,
\label{D+}
\end{equation}
and hence $|\Omega^*_{-}(0)\rangle = |\Omega^*_{-}\rangle $, $|\Omega^*_{-}(\infty)\rangle = |\Omega_{-}\rangle $.
Therefore, the time-dependent state of the pointer 
\begin{equation}
\hat{\rho}_{\mathcal{P}}(t) = |c_{-}|^2|D_{-}(t)\rangle\langle D_{-}(t)| + |c_{+}|^2|D\rangle\langle D|
\label{pointer}
\end{equation}
asymptotically tends to the final state which satisfies the Born rule of the standard measurement theory
\begin{equation}
\hat{\rho}_{\mathcal{P}}(\infty) = |c_{-}|^2|-D\rangle\langle -D| + |c_{+}|^2|D\rangle\langle D|
\label{pointer_fin}
\end{equation}
with macroscopically distinguishable pointer states.
\par
The computation of above shows how the result of the initial quantum CNOT gate is stabilized and written on the stable semiclassical information carrier. The process of stabilization is strongly irreversible and accompanied by the dissipation of the energy $\bar{W}$. The final result is the irreversible classical CNOT gate performed on the pointer states distinguishable with the error 
\begin{equation}
\epsilon = e^{-4D^2} = \exp\Bigl\{-\frac{\bar{W}}{\frac{1}{2}\omega_0}\Bigr\},
\label{error}
\end{equation}
where \emph{zero point energy} $\frac{1}{2}\omega_0$ characterizes quantum fluctuations of the pointer.
\section{Finite temperature case}
\par
In this section I present the generalization of the model discussed above to the case of a more realistic and more interesting finite-temperature environment.
\par
The first difference between the finite and zero temperature case is the form of spectral densities
\begin{equation}
G^{(T)}_j(\omega)= \int_{-\infty}^{+\infty} e^{i\omega t}\langle \hat{F}_j(t)\hat{F}_j\rangle_{T} dt \geq 0 ,
\label{spectral_bath_T}
\end{equation}
where now $\langle \cdot \rangle_{T}$ denotes the average with respect to Gibbs state at the temperature $T$ what implies the following Kubo-Martin-Schwinger (KMS) relation
\begin{equation}
G^{(T)}_j(-\omega)= e^{-\omega/T} G_j(\omega).
\label{KMS}
\end{equation}
Again, first disregarding the environmental tunneling, one obtains using standard derivation the following quantum Markovian master equation for the density matrix of SOS 
\begin{eqnarray}
\frac{d\hat{\rho}}{dt} = -i[\hat{H} , \hat{\rho}]  + \frac{1}{2}\gamma\bigl([\hat{b}, \hat{\rho}\hat{b}^{\dagger}] + [\hat{b} \hat{\rho}, \hat{b}^{\dagger}]\bigr)+ \frac{1}{2}\gamma e^{-\omega_0/T}\bigl([\hat{b}^{\dagger}, \hat{\rho}\hat{b}] + [\hat{b}^{\dagger} \hat{\rho}, \hat{b}]\bigr) -\frac{1}{2}\Gamma [\hat{\tau}^3,[\hat{\tau}^3, \hat{\rho}]].
\label{MME_T}
\end{eqnarray}
where the dissipation rate $\gamma = G^{(T)}_o(\omega_0)$, and the pure decoherence $\Gamma = 4D^2 G^{(T)}_o(0) + G^{(T)}_3(0)$.
Similarly to \eqref{MME} the master equation \eqref{MME_T} is exactly solvable as in the dressed states picture the evolution is given by a product of the standard linearly damped and pumped harmonic oscillator dynamics and the pure decoherence dynamics for the spin. The exact solution can be given in different terms, e.g. by Gaussian propagators in position or momentum representations, P, Q or Wigner representations, etc. I am not going to discuss the explicit solutions for arbitrary initial states but concentrate only on the special cases relevant for the description of the measurement process.
\par
The two stable ground states of the SOS are replaced now by a pair of \emph{biased Gibbs states}
\begin{equation}
\hat\rho^{(\pm)} =  \bigl(1-e^{-\omega_0/T}\bigr)|\pm\rangle\langle\pm|\,e^{-\frac{\omega_0}{T}\hat{b}^{\dagger}\hat{b}} = \bigl(1-e^{-\omega_0/T}\bigr)|\pm\rangle\langle\pm|\,e^{-\frac{\omega_0}{T}(\hat{a}^{\dagger}\mp D)(\hat{a}\mp D)}
\label{gibbs_biased}
\end{equation}
and any initial SOS state $\hat{\rho}$ tends asymptotically to the mixture $p_{+}\hat\rho^{(+)} + p_{-}\hat\rho^{(-)}$  with $p_{\pm}$ given by \eqref{prob_in}. The corresponding stable pointer (oscillator) states are now mixed states but well-localized and distinguishable (for $D>>1$) and given by
\begin{equation}
\hat\rho^{(\pm)}_{\mathcal{P}} =  \bigl(1-e^{-\omega_0/T}\bigr)e^{-\frac{\omega_0}{T}(\hat{a}^{\dagger}\mp D)(\hat{a}\mp D)}.
\label{pointer_T}
\end{equation}
Their overlap defined as the transition probability (see \cite{Scutaru:1998} for the derivation) reads
\begin{equation}
\epsilon = \mathrm{Tr}\Bigl(\sqrt{\sqrt{\hat\rho_{\mathcal{P}}^{(+)}}\hat\rho_{\mathcal{P}}^{(-)}\sqrt{\hat\rho_{\mathcal{P}}^{(+)}}}\Bigr) =  \exp\Bigl\{-4 D^2\tanh\bigl(\frac{\omega_0}{2T}\bigr)\Bigr\}
\label{trans_prob}
\end{equation}
and describes probability of error in the process of their discrimination (measurement error). The formula interpolates between the zero-temperature value given by \eqref{trans_0} and the  high temperature one determined by the Boltzmann factor, and can be rewritten as
\begin{equation}
\epsilon =  \exp\Bigl\{-\frac{\bar{W}}{\Theta}\Bigr\} , \quad \Theta= \frac{\omega_0}{e^{\omega_0/T} -1} + \frac{\omega_0}{2}.
\label{highT}
\end{equation}
Here $\Theta \equiv \Theta[T,\omega_0]$ is the average quantum oscillator energy which in the semiclassical regime, i.e. for $\frac{\omega_0}{T} << 1$, is equal to the temperature, $\Theta\simeq T$, while for $\frac{\omega_0}{T} >> 1$, is equal to the zero point energy $\Theta\simeq \frac{1}{2}\omega_0$. One can call $\Theta$ \emph{noise temperature} because  it characterizes  noise acting on the pointer which has a \emph{thermal fluctuation} component and a \emph{quantum fluctuation} one.
\par
The formula \eqref{highT} implies one of the main results of this paper:\\

\emph{The minimal work needed to encode a bit of information with an error probability $\epsilon$ under the influence of combined thermal and quantum noise at the noise temperature $\Theta$ is given by}
\begin{equation}
\bar{W} =  \Theta \ln{\frac{1}{\epsilon}}.
\label{minwork}
\end{equation}
The minimal work is always larger than Landauer's $T\ln 2$ ($\epsilon \leq \frac{1}{2}$) and does not vanish for $T\to 0$, because it includes also quantum fluctuations.
\par
Obviously, the same amount of work is needed to \emph{reset} the one-bit memory to a fixed reference state. However reseting of the memory is not necessary, because the pointer states are distinguishable and one can use a \emph{change} of the pointer position as a signal carrying a bit of information. One should notice the difference between the reseting process and the \emph{forgeting} process represented here by the dissipative tunneling due to noise.
\par
The measurement scheme is exactly the same as for the zero-temperature bath. The initial ground state of SOS $|\Omega_{+}\rangle$ is replaced by the biased Gibbs state $\hat\rho^{(+)}$ and the total initial state of  $\mathcal{O}$ and SOS reads
\begin{equation}
\hat{\rho}_{\mathrm{tot}}= \Bigl[|c_{-}|^2|\phi_{-}\rangle\langle \phi_{-}| + |c_{+}|^2|\phi_{+}\rangle\langle \phi_{+}|
+ \bigl(c_{-}\overline{c_{+}}|\phi_{-}\rangle\langle \phi_{+}|+\mathrm{h.c.}\bigr)\Bigr]\hat\rho^{(+)}.
\label{tot_state_1}
\end{equation}
The state just after measurement is given by
\begin{equation}
\hat{\rho}_{\mathrm{tot}}(t_M)= |c_{-}|^2|\phi_{-}\rangle\langle \phi_{-}|{\hat\rho^{(-)}}_* + |c_{+}|^2|\phi_{+}\rangle\langle \phi_{+}|\hat\rho^{(+)}
+ \bigl(c_{-}\overline{c_{+}}|\phi_{-}\rangle\langle \phi_{+}|\hat{\sigma}^1 \hat\rho^{(+)}+\mathrm{h.c.}\bigr)
\label{tot_state2}
\end{equation}
where the excited biased Gibbs state ${\hat\rho^{(\pm)}}_*$  is defined as (compare with \eqref{gibbs_biased})
\begin{equation}
\hat\rho^{(\pm)}_* = \bigl(1-e^{-\omega_0/T}\bigr)|\pm\rangle\langle\pm|e^{-\frac{\omega_0}{T}(\hat{a}^{\dagger}\pm D)(\hat{a}\pm D)}.
\label{gibbs_ex}
\end{equation}
Due to the fact that $\mathrm{Tr}(\hat{\sigma}^1 \hat\rho^{(\pm)})= 0$ the state of $\mathcal{O}$ after the measurement is again given by \eqref{Ostate_post} and for $t > t_M$ SOS and $\mathcal{O}$ evolve independently. 
Neglecting again the tunneling process one can compute the reduced density matrix of the pointer which possesses a similar structure to \eqref{pointer} 
\begin{equation}
\nonumber
\hat{\rho}_{\mathcal{P}}(t) = |c_{-}|^2\,\hat\rho_{\mathcal{P}}^{(+)}(t) + |c_{+}|^2\,\hat\rho_{\mathcal{P}}^{(+)} 
\label{SOS_state1}
\end{equation}
where
\begin{equation}
\hat\rho^{(+)}(t)= \bigl(1-e^{-\omega_0/T}\bigr)\,e^{-\frac{\omega_0}{T}(\hat{a}^{\dagger}-D_{-}(t))(\hat{a}- D_{-}(t))}, \quad    D_{-}(t) = D \bigl(2 e^{-(i\omega_0 +\frac{\gamma}{2})t}-1\bigr) .
\label{rho+}
\end{equation}
Because $\hat\rho_{\mathcal{P}}^{(+)}(0)= \hat\rho^{(+)}_{\mathcal{P}}$, $\hat\rho_{\mathcal{P}}^{(+)}(\infty)= \hat\rho^{(-)}_{\mathcal{P}}$ the final state of the pointer satisfies the Born rule of the standard measurement theory
\begin{equation}
\hat{\rho}_{\mathcal{P}}(\infty) = |c_{-}|^2\,\hat\rho^{(-)}_{\mathcal{P}} + |c_{+}|^2\,\hat\rho^{(+)}_{\mathcal{P}}.
\label{pointer1}
\end{equation}
Finally, the slow tunneling process should be added. Using the same assumptions as for the zero-temperature case we can derive the analog of the formula \eqref{tunrate} replacing the ground state by the biased Gibbs one
\begin{equation}
\Gamma_{tun} = \frac{1}{2}\mathrm{Tr}\Bigl(\hat{\tau}^3 \frac{d\hat{\rho}}{dt}|_{t=0} \Bigr)=\frac{1}{2}\mathrm{Tr}\Bigl(\hat{\tau}^3\mathcal{L}^{(1)}\hat\rho^{(+)} \Bigr)= \frac{1}{2}G_1(0)\mathrm{Tr}\bigl((\hat{B}_0)^2\hat\rho^{(+)}\bigr)
\label{tunrate_temp}
\end{equation}
The rather lengthly calculations presented in the Appendix lead to the following estimation of the minimal tunneling rate
\begin{equation}
\Gamma_{tun} \simeq \frac{1}{2}G_1(0)e^{-\frac{\bar{W}}{\Theta}}.
\label{tunrate_temp1}
\end{equation}
\par
One can now summarize the description of all stages of the measurement process 
taking place on well-separated time-scales:\\

1) Fast unitary preparation of the entangled state of $\mathcal{O}$ and the spin-$1/2$ interface  $c_{-}|\phi_{-}\rangle|-\rangle + c_{+}|\phi_{+}\rangle |+\rangle$ is performed. The process takes (in principle) an arbitrarily short \emph{measurement time} $t_M$  and needs $|c_{-}|^2 4\omega_0 D^2$ of work to execute the quantum CNOT gate.
The reduced state of $\mathcal{O}$, just after $t_M$, is a standard post-measurement mixed state $|c_{-}|^2|\phi_{-}\rangle\langle \phi_{-}| + |c_{+}|^2|\phi_{+}\rangle\langle \phi_{+}|$
which evolves subsequently according to the  Hamiltonian of $\mathcal{O}$.\\

2) \emph{Dequantization} irreversible process, due to SOS-environment coupling  which kills the quantum coherences between the emerging Schroedinger cat states of SOS. It takes short \emph{dequantization time} $t_D\sim\frac{1}{D^2}$.\\

3) The essentially classical (conditional) evolution of the pointer  well-localized Gaussian state along the classical damped harmonic oscillator trajectory from the initial zero momentum state localized at $x=D$ to a final pointer state localized at $x=-D$. Here, the characteristic relaxation time does not depend on $D$ and is equal to $t_R = \frac{1}{\gamma}$ - the \emph{recording time}.\\

4) A very slow \emph{erasure process} of the measurement result which takes place on the \emph{memory time} scale
$t_E \sim e^{\frac{\bar{W}}{\Theta}}\sim \frac{1}{\epsilon}$.
\section{Quantum Szilard engine and its efficiency}
The original Szilard engine is based on the Maxwell's set-up, but with only a single gas particle in a box. If after the measurement one knows which half of the box is occupied by the particle (single bit of information), one can  close a piston unopposed into the empty half of the box, and then extract $ T \ln 2 $ of  work by the isothermal expansion. \\
The quantum analog of the Szilard engine, discussed for example in \cite{Alicki:2004}, consists of a TLS governed by the time-dependent Hamiltonian 
\begin{equation}
H(t)= \frac{E_{0}}{2}\bigl(f(t)^2 \hat{I} - f(t)\hat{\sigma}^3 \bigr)
\label{Szil_ham}
\end{equation}
with the external control $|f(t)| \leq 1$ and $E_{0}>0$. The weak coupling to the heat bath at the temperature $T$ can be switched on and off.\\
The cyclic process of extracting work from a bath using a bit of information consists of the following steps :\\

i) For the initial time $t_0$ the Hamiltonian is trivial, i.e. $f(t_0)=0$ and the TLS,  coupled to the bath, is at the corresponding thermal equilibrium state $\hat{\rho}(t_0)= \frac{1}{2}\hat{I}$.\\

ii) The coupling to the bath is switched off and a measurement on the TLS in the basis of $\hat{\sigma}^3$ is performed giving  the outcome 
$s= \pm 1$ with the corresponding projected post-measurement state $\hat{\rho }(t_1)= |s\rangle\langle s|$.\\

iii) A fast (in comparison to the thermal relaxation time) change of the external field from the value $f(t_1)=f(t_0)=0$ to the value $f(t_2) = s$ is performed  producing the Hamiltonian  $H(t_2) = (s E/2)(s\hat{I}-\hat{\sigma}^3 )$ which increases the energy of the state $|-s\rangle$ by $E_{0}$ and  does not change the energy of
the state $|s\rangle$. As the actually occupied state does not change its energy no work is performed during this stage.
\\

iv) The coupling of TLS to the baths is switched on and the external field is slowly reduced (again slowly in comparison to the thermal relaxation time) from the value  $f(t_2) = s$  to the value $f(t_3) = 0$.\\
\par
One can  compute the balance of work $W(t)$ and  heat $Q(t)$ supplied to TLS, and  its internal energy $E(t) $ during the full cycle $t_0\to t_1\to t_2$
using the standard definitions (discussed in the quantum context in \cite{Alicki:1979, Alicki:2004}) 
\begin{equation}
E=\mathrm{Tr }(\rho H),\quad dW = \mathrm{Tr }(\rho\, dH), \quad dQ =\mathrm{Tr } (d\rho\, H).
\label{EWQ}
\end{equation}
Notice, that in the case of a \emph{perfect measurement}, only in the step iv)  work
is adiabatically extracted from the bath. Due to the slow change of the Hamiltonian one can assume that at any moment  the TLS is in the thermal equilibrium state with respect to the temporal  Hamiltonian \eqref{Szil_ham} at the bath temperature. This state is given by the Gibbs expression 
\begin{equation}
\rho (t) = \frac{ \exp\{-\frac{H(t)}{T}\}}{\mathrm{Tr} \exp\{-\frac{H(t)}{T}\}},\quad t_2 \leq t\leq t_3
\label{Gibbs_temp}
\end{equation}
and according to \eqref{EWQ} work performed by the Szilard engine during the whole cycle is given by
\begin{equation}
W_{SE}(E_0) = -\int_{t_2}^{t_3}\mathrm{Tr}\bigl(\rho (t)\frac{dH(t)}{dt}\bigr) dt =  T\bigl[\ln 2 - \ln\bigl(e^{-\frac{E_0}{T}}+ 1\bigr)\bigr]
\label{SE_work}
\end{equation}
which for $E_0\mapsto\infty$ reaches the well-known value $T \ln 2$ - \emph{maximal work which can be extracted from a heat bath using a bit of information}. 
\subsection{Landauer's principle}
The possibility of extracting work from a single heat bath in a cyclic process apparently violates the Second Law of Thermodynamics. To avoid this conflict with the Second Law
one has to conclude that the following Landauer's principle for measurement  holds :\\
\emph{ A completion of a binary measurement, including reseting of a measuring device needs at least $T\ln 2 $ of work.}\\
\par
Notice that the arguments of above doe not apply to the recognition cost of the stable pointer state. Namely, in the step iv the relaxation  time should be much faster that the time devoted to extract work, what is not the case for stable pointer states.
\par
As argued in \cite{Bennett:2003} the amount of work (at least $T\ln 2$) needed to perform a binary measurement is actually used to erase a bit of  information in a memory of a measuring device. More generally, one claims that  energy cost of any irreversible elementary gate is also of the same order. However, the existing arguments based on microscopic models of erasure and  entropy-energy balance (see e.g.\cite{Piech}) are in my opinion not convincing and have been criticized in \cite{Alicki:2012}. On the other hand the formula \eqref{minwork} suggests the following principle:\\

\emph{The minimal work needed to perform an elementary gate  on a protected  information carrier is of the order of $\Theta \ln \frac{1}{\epsilon}$, where $\epsilon$ is the probability of readout error and $\Theta$ is the effective noise temperature. Moreover, the life-time of protected information scales like $\frac{1}{\epsilon}$.} 

\subsection{Szilard engine with faulty measurement}

The maximal work extracted by the Szilard engine has been computed under the assumption of perfect measurement. It has been assumed also that the efficiency of this process can reach one, i.e. whole work invested in the erasure of a bit can be, in principle, extracted.
\par
If the measurement yields a correct result with the probability $1-\epsilon$, $\epsilon\in[0, 1/2]$, then in the step iii) of the cycle the amount of work $E_0$ is supplied by the external field with the probability $\epsilon$. Therefore, the net extracted work is given by
\begin{equation}
W_{SE}(E_0;\epsilon) = T\bigl[\ln 2 - \ln\bigl(e^{-\frac{E_0}{ T}}+ 1\bigr)\bigr]- \epsilon E_0.
\label{SE_work1}
\end{equation}
Maximizing $W_{SE}(E_0;\epsilon)$ with respect to $E_0$ one obtains
\begin{equation}
W_{SE}[\epsilon] = \max_{E_0}W_{SE}(E_0;\epsilon) =  T\bigl[\ln 2 - S(\epsilon) \bigr].
\label{SE_work2}
\end{equation}
where $S(\epsilon) = -\epsilon \ln\epsilon - (1-\epsilon) \ln(1-\epsilon)$ corresponds to the entropic uncertainty of the measurement. The efficiency of the Szilard engine can be defined as the ratio of the extracted work to the minimal work $\bar{W}$  needed to perform the measurement and satisfies the following bound obtained numerically
\begin{equation}
\eta[\epsilon] = \frac{W_{SE}[\epsilon]}{\bar{W}} = \frac{T}{\Theta}\frac{\ln 2 - S(\epsilon)}{-\ln \epsilon}\leq \eta[\bar{\epsilon}]= 0.17 \frac{T}{\Theta}< 0.17.
\label{SE_eff}
\end{equation}
The maximal efficiency of Szilard engine is obtained for the value of error $\bar{\epsilon}= 0.06$.
\section{Concluding remarks}
The formulas derived above  provide relations between stability, accuracy and thermodynamical cost for a single
gate performed on the protected single-bit information carrier. These results allow also to address a more general question: \emph{What is the minimal work needed to perform an algorithm which consists of $N$ elementary logical steps?}
\par
To find a proper estimation I make the  assumption that the time $\tau_{gate}$ needed to complete an elementary gate  is of the order of \emph{recording time} $t_R$  and is  given by $\tau_{gate} \simeq \frac{1}{\gamma}$, where $\gamma$ denotes the relaxation rate appearing in the eq. \eqref{MME_T}. Another important parameter  is the ratio of the relaxation rate $\gamma $ to the (decoupled) spin decoherence rate $\frac{1}{2}G_1(0)$ 
\begin{equation}
\kappa = \frac{2\gamma }{G_1(0)}.
\label{SE_eff}
\end{equation}
Obviously, the value of $\kappa$, as well as the frequency scale $\omega_0$  depend on the technology used to implement the information carrier. 
\par
Under those assumption the time needed to perform the algorithm is equal to $\tau_{N} = \frac{N}{\gamma}$ and the probability of error due to the \emph{forgeting process} characterized by the tunneling time $t_E >> \tau_{N}$  can be estimated as 
\begin{equation}
\delta = \frac{\tau_{N}}{t_{E}} \simeq \frac{N \gamma^{-1}}{\bigl[{\frac{1}{2}G_1(0)}\bigr]^{-1}e^{\frac{\bar{W}}{\Theta} }} = \frac{1}{\kappa}N e^{-\frac{\bar{W}}{\Theta} }.
\label{comp_error}
\end{equation}
Therefore, the minimal amount of work necessary to perform the algorithm is equal to $ \bar{W}_N = N \bar{W}$ and can be estimated as
\begin{equation}
\bar{W}_N \simeq \Theta  N\bigl(  \ln N +  \ln \frac{1}{\delta}+  \ln \frac{1}{\kappa}\bigr) .
\label{comp_error1}
\end{equation}
One can notice the differences between the formula \eqref{comp_error1} and the prediction based on the Landauer principle $\bar{W}_N^{(L)} \simeq T  N \ln 2$. The work is non-additive with respect to the algorithm size $N$,  depends on the assumed failure probability $\delta$ and the parameter $\kappa$ determined by the implementation, and does not vanish for $T\to 0$. For example, a modern supercomputer performing $10^{16}$ logical gates per second and working for a day executes an algorithm with $N \simeq 10^{21}$.  As a reasonable benchmark for the parameter $\kappa$ one can take the minimal ratio of relaxation times $\frac{T_2}{T_1} \simeq 10^{-8}$ in NMR experiments, or the ratio of natural line width to the collisionally broaden one in atomic spectroscopy which can reach the comparable values. Then, under the  assumption that $\delta\times \kappa >> 10^{-21}$, and at the room temperature $\Theta\simeq 300 K$ the total minimal work  $\bar{W}_N \simeq 10^2 J$ what is still much less than the actual energy consumption, which is of the order of $10^{10} J$.
\par
Another prediction of this model is the conflict between reversibility and stability of information processing (see also \cite{Alicki:2012}). Namely, the more stable are information carriers the more work must be invested in a logical gate. This work is subsequently dissipated making the gates strongly irreversible. The irreversibility (nonunitarity) of information processing does not harm classical computations but can put the limits on large scale quantum ones.\\

\textbf{Acknowledgements}\\
The authors thanks Krzysztof Szczygielski for the assistance and acknowledges the support by the FNP TEAM project cofinanced by EU Regional Development Fund.\\

\section{Appendix}
In this Appendix the detailed derivation of the initial tunneling rate for the finite temperature case is presented.
To construct the semigroup generator $\mathcal{L}^{(1)}$ which is responsible for the dissipative tunneling process we use the Fourier decomposition of the interaction operator $\hat{\sigma}^1(t)$ (compare \eqref{Fourier_3})
\begin{eqnarray}
\label{fourier_1}
\hat{\sigma}^1(t)&=& e^{i\hat{H}_S t} \hat{\sigma}^1 e^{-i\hat{H}_S t}  =  e^{2D( e^{-i\omega_0 t}\hat{b}- e^{i\omega_0 t}\hat{b}^{\dagger})}\hat{\tau}^{+} + e^{-2D( e^{-i\omega_0 t}\hat{b}- e^{i\omega_0 t}\hat{b}^{\dagger})}\hat{\tau}^{-}\\
\nonumber
& = & \hat{W}(t) \hat{\tau}^{+} +\hat{W}^{\dagger}(t) \hat{\tau}^{-}\\
\nonumber
& = & \sum_{m\in\mathbf{Z}} e^{im\omega_0 t}\bigl(\hat{W}_m \hat{\tau}^{+} +\hat{W}^{\dagger}_{-m} \hat{\tau}^{-}\bigr)
\end{eqnarray}
Applying now the standard construction of the weak-coupling regime generator \cite{Davies:1974} one obtains its following form
\begin{equation}
\mathcal{L}^{(1)}\hat{\rho} = \frac{1}{2}\sum_{m\in\mathbf{Z}}G_1(m\omega_0 )\Bigl\{\bigl[(\hat{W}_{-m} \hat{\tau}^{+} +\hat{W}^{\dagger}_{m}\hat{\tau}^{-})\hat{\rho},(\hat{W}_m \hat{\tau}^{+} +\hat{W}^{\dagger}_{-m})\hat{\tau}^{-}\bigr] + \mathrm{h.c.}\Bigr\}
\label{L_1}
\end{equation}
The initial rate of the probability flow out of the fixed biased Gibbs state $\hat\rho^{(+)}$ which characterizes  stability of the pointer states reads
\begin{equation}
\Gamma_{tun} = \frac{1}{2}\mathrm{Tr}\Bigl(\hat{\tau}^3 \frac{d\hat{\rho}}{dt}|_{t=0} \Bigr)=\frac{1}{2}\mathrm{Tr}\Bigl(\hat{\tau}^3\mathcal{L}^{(1)}\hat\rho^{(+)} \Bigr).
\label{tunrate_A}
\end{equation}
After staightforward computation one obtains 
\begin{equation}
\Gamma_{tun} = \frac{1}{2}\sum_{m\in\mathbf{Z}}G_1(m\omega_0 )(1 - e^{-\frac{\omega_0}{T}})\mathrm{Tr}\Bigl(e^{-\frac{\omega_0 \hat{a}^{\dagger}\hat{a}}{T}} \hat{V}_m  \hat{V}^{\dagger}_m \Bigr)
\label{tunrate_A1}
\end{equation}
where
\begin{equation}
\hat{V}(t) =   e^{2D( e^{-i\omega_0 t}\hat{a}- e^{i\omega_0 t}\hat{a}^{\dagger})}
 = \sum_{m\in\mathbf{Z}} e^{im\omega_0 t} \hat{V}_m .
\label{Voperator}
\end{equation}
Typically, for systems in the semiclassical regime, i.e. for $D >> 1$,  pure decoherence effects dominate over the dissipative ones. It means that in the formula \eqref{tunrate_A1} the term with $m=0$ is a leading factor and from now on the minimal value of tunneling rate will be estimated as follows ($\hat{V}_0 = \hat{V}^{\dagger}_0 $)
\begin{equation}
\Gamma_{tun} \simeq \frac{1}{2}G_1(0)(1 - e^{-\frac{\omega_0}{T}})\mathrm{Tr}\Bigl(e^{-\frac{\omega_0 \hat{a}^{\dagger}\hat{a}}{T}} \hat{V}_0 ^2 \Bigr) = \frac{1}{2}G_1(0)\langle  \hat{V}_0 ^2 \rangle_T.
\label{tunrate_A2}
\end{equation}
where $\langle \cdots \rangle_T $ denotes thermal average for the harmonic oscillator. To compute $\langle\hat{V}_0 ^2 \rangle_T$ one can use the  identities which follow from \eqref{Voperator} and periodicity of $\hat{V}(t)$
\begin{equation}
\hat{V}_0= \frac{1}{2\pi}\int_0^{2\pi}\hat{V}(t)dt = \lim_{a\to\infty} \frac{1}{2a}\int_0^{2a}\hat{V}(t)dt ,
\label{Voperator1}
\end{equation}
\begin{equation}
\langle\hat{V}_0^2\rangle_T =  \lim_{a\to\infty} \frac{1}{4a^2}\int_0^{2a}\int_0^{2a}\langle\hat{V}(t)\hat{V}^{\dagger}(s)\rangle_T \,dt\,ds  
 =  \lim_{a\to\infty} \frac{1}{2a}\int_0^{2a}\langle\hat{V}(t)\hat{V}^{\dagger}(0)\rangle_T \,dt =\frac{1}{2\pi}\int_0^{2\pi}\langle\hat{V}(t)\hat{V}^{\dagger}(0)\rangle_T \,dt.
\label{Voperator2}
\end{equation}
Both operators $\hat{V}(t)$ and $\hat{V}^{\dagger}(0)$ are particular examples of Weyl unitaries parametrized by complex numbers $\alpha$ and defined as $\hat{W}(\alpha)= \exp\{\alpha \hat{a} - \bar{\alpha}\hat{a}^{\dagger}\}$. Using their composition law $\hat{W}(\alpha)\hat{W}(\beta) = \exp\{\frac{1}{2}(\bar{\alpha}\beta - \alpha\bar{\beta}\}\hat{W}(\alpha + \beta)$ and the formula for the thermal average
\begin{equation}
\langle\hat{W}(\alpha)\rangle_T = \exp\Bigl\{-\frac{|\alpha|^2}{2\bigl(1- e^{-\frac{\omega_0}{T}} \bigr)}\Bigr\}
\label{thermal_W}
\end{equation}
one can compute
\begin{eqnarray}
\langle\hat{V}_0^2\rangle_T &= &\exp\Bigl\{-\frac{4D^2}{1- e^{-\frac{\omega_0}{T}}}\Bigr\} \int_0^{2\pi} \exp\Bigl\{{4D^2}\Bigl[\frac{\cos x}{1- e^{-\frac{\omega_0}{T}}} -i \sin x\Bigr]\Bigr\} dx \\
\nonumber
&= &\exp\Bigl\{-\frac{4D^2}{1- e^{-\frac{\omega_0}{T}}}\Bigr\}\, \mathrm{I}_0 \Bigl(4D^2 \sqrt{(1- e^{-\frac{\omega_0}{T}})^{-2} - 1}\Bigr) ,
\label{Voperator}
\end{eqnarray}
where $\mathrm{I}_0$ is a modified Bessel function. Using the fact that $\mathrm{I}_0(0) =1$  and for $x >> 1$  $\mathrm{I}_0(x) \simeq \frac{e^x}{\sqrt{2\pi x}}$  we can take into account the leading exponential term to get the following approximation
\begin{equation}
\Gamma_{tun} \simeq \frac{1}{2}G_1(0)e^{-\frac{\bar{W}}{\Theta'}}
\label{tunrate_A3}
\end{equation}
where again as in \eqref{highT} $\bar{W} = 2D^2 \omega_0$ and
\begin{equation}
{\Theta'}= \frac{\omega_0}{2}\frac{1- e^{-\frac{\omega_0}{T}}}{1-\sqrt{1 - (1- e^{\frac{-\omega_0}{T}})^2}} .
\label{new_temperature}
\end{equation}
The new effective noise temperature $\Theta'$ is different from that defined in \eqref{highT}, however the difference is small. Their asymptotic values for low and high temperatures are the same and the maximal difference between $\Theta$ and $\Theta'$ reaches  $30\%$ in the crossover region $\omega_0 \simeq T$. Therefore, one can put in the estimation $\Theta '\simeq \Theta $ and hence, finally
\begin{equation}
\Gamma_{tun} \simeq \frac{1}{2}G_1(0)e^{-\frac{\bar{W}}{\Theta}}\simeq \frac{1}{2}G_1(0) \frac{1}{\epsilon} .
\label{tunrate_A4}
\end{equation}

\end{document}